# Possible further enhancement of the supercurrent carrying ability in $MgB_2$ using chemical doping: insights from the substantial enhancement of in-field $J_c$ in $MgB_2$ using liquid $SiCl_4$ as a dopant

Xiao-Lin Wang[1], Shi-xue Dou[1], M.S.A. Hossain[1], Zhen-xiang Cheng[1], Xiao-zhou Liao[2], S. R. Ghorbani[1,3]

[1]*Institute for Superconducting and Electronic Materials, University of Wollongong, Wollongong, NSW 2522, Australia*

[2]*School of Aerospace, Mechanical & Mechatronic Engineering, Building J07, the University of Sydney, NSW 2006, Australia*

[3]*Department of Physics, Tarbiat Moallem University of Sabzevar, P.O. Box, 397, Sabzevar, Iran*

In this work, we present the following important results: 1) We introduce a new Si source, liquid $SiCl_4$, which is free of C, to significantly enhance the irreversibility field ($H_{irr}$), the upper critical field ($H_{c2}$), and the critical current density ($J_c$), with little reduction in the critical temperature ($T_c$). 2) Although Si can not incorporate into the crystal lattice, we found a reduction in the a-axis lattice parameter, to the same extent as for carbon doping. 3) The $SiCl_4$ treated $MgB_2$ shows much higher $J_c$ with superior field dependence above 20 K than undoepd $MgB_2$ and $MgB_2$ doped with various carbon sources. 3) We provide an alternative interpretation for the reduction of the *a* lattice parameter in C- and non-C doped $MgB_2$. 4). We introduce a new parameter, RHH ($H_{c2}/H_{irr}$), which can clearly reflect the degree of flux pinning enhancement, providing us with guidance for further enhancing $J_c$. 5) We have found that spatial variation in the charge carrier mean free path is responsible for the flux pinning mechanism in the $SiCl_4$ treated $MgB_2$ with large in-field $J_c$.

Magnesium diboride superconductor ($MgB_2$)[1], with a much higher superconducting transition temperature *($T_c$)* of 40 K and lower cost than conventional low temperature superconductors ($T_c$ < 25 K), has great potential for large-scale and microelectronic applications at temperatures far above that of liquid helium (4.2 K). For practical applications that require carrying large supercurrents in the presence of magnetic field, improvement in the critical current density ($J_c$) has been the key research topic for $MgB_2$, although the weak link problem is almost negligible[2] compared to the high $T_c$ cuprate superconductors (HTS). The irreversibility field, $H_{irr}$, is the maximum field at which $MgB_2$ lose its supercurrent. There are three ways to increase $H_{irr}$: 1) increasing $T_c$; 2) increasing the upper critical field, $H_{c2}$; and 3) introducing effective pinning centers to enhance the flux pinning if $H_{c2}$ is constant. However, the $T_c$ of $MgB_2$ always decreases for various dopants that can substitute into either Mg or B sites and can be significantly reduced to very low temperatures far below 40 K[3], limiting applications at higher temperatures above 20 K. The enhancement of $H_{c2}$ has been very successful via various chemical doping, such as nano-SiC, Si and C[4-9] carbohydrates[10] containing C, H and O and liquid additive, silicone oil containg Si, C, H and O[11]..

So far, it has been widely accepted that, among all the chemical dopants, carbon is the most successful doping element for significantly enhancing $H_{c2}$, depending on the carbon doping level. The reason is that the B sites are partially occupied by carbon and cause strong $\sigma$ band electron scattering, which enhances $H_{c2}$ significantly[12]. However, the disadvantage of carbon doping is that it leads to a significant drop in $T_c$ from 39 K down to near 10 K, for carbon content up to 20 %.[13]. This means that a small amount of C can significantly increase the $H_{c2}$, but only with large sacrifices in $T_c$. As a consequence, the significant reduction in $T_c$ due to C doping bars $MgB_2$ from applications above 20 K, which is the critical temperature at which the $MgB_2$ is most likely to be useful in liquid helium free magnetic resonance imaging (MRI) devices. As for the third approach, if both the $T_c$ and $H_{c2}$ are constant, the $H_{irr}$ can be moved up by introducing effective pinning centers. This is the case where there is enhancement of effective flux pinning.

Furthermore, it should be pointed out that there are three issues to do with pinning enhancement that have to be clarified for $MgB_2$ doping with C using various carbon sources. 1) What role do C and the residual C play in flux pinning?; 2) Does C really occupy B sites in the case of samples made at low sintering temperatures?; 3). Can we use empirical rules based on C-doped $MgB_2$ single crystals or the reduction in the *a* lattice parameter to estimate the carbon concentration in the samples doped using various carbon sources?; 4) Can we find another dopant that is capable of enhancing $J_c$ as much as C does, but without degradation in $T_c$?; 5) What has really been improved for all the $MgB_2$ in all the chemical doping studies? Or, what are the effective approaches to judge whether or not effective flux pinning has been introduced into $MgB_2$? The above questions are the motivations behind this work, and all these issues are discussed.

The RRR ratio, i.e the ratio of the resistivity at 300 K to that at $T_c$ has been widely used as a parameter to reflect the degree of electron scattering, which is believed to be related to flux pinning. However, in most cases, the RRR ratios only give an indication of electron scattering and may have no correlation with the degree of pinning. Here, we introduce another new ratio, RHH, the ratio of $H_{c2}(T)$ to $H_{irr}(T)$. As shown in Fig.1(a) and (b), for a sample with a fixed $H_{c2}$, the position of the irreversibility line reflects the degree of flux pinning. The smaller the distance between the $H_{irr}$ and $H_{c2}$ lines, the stronger the flux pinning, while a large distance means weaker pinning. Generally, $H_{irr}$ rises in proportion to any increase in $H_{c2}$, because the same defects that are effective for electron scattering also tend to pin flux. In the case of $MgB_2$ with a higher $H_{c2}$, however (see Fig. 1(c) and (d)), the $H_{irr}$ line may not necessarily increase in proportion to the increase in $H_{c2}$. In this case, flux pinning is not enhanced, even though the absolute value of $H_{irr}$ may be higher than that of $MgB_2$ with lower $H_{c2}$. It is obvious that if the RHH value is close to 1, the flux pinning, and thus $H_{irr}$, has reached its limit. In other words, if effective pinning centers are present in a $MgB_2$ sample, the RHH should be as close to 1 as possible. If RHH is greater than 2, that means an $H_{irr}$ that is only half as great as $H_{c2}$. If the RHH >> 2, one can say that the flux pinning is very weak. This new ratio will be used to judge if effective flux pinning is present in $MgB_2$ samples, allowing comparisons to be made in this work.

## Experimental

The differences between the $SiCl_4$ and all the other above-mentioned doping chemicals are as follows: 1). $SiCl_4$ does not produce C at all, while all the other $J_c$ enhancing dopants contain carbon. 2). In addition to silicone oil[11], $SiCl_4$ is a typical inorganic liquid containing Si. It is widely used to produce Si nanowires as one of the starting precursors. Therefore, it is an ideal inorganic liquid for doping $MgB_2$. 3). As it is a liquid under ambient condition, it offers a great advantage in terms of very homogeneous mixing with solid state particles such as boron, as compared to the problem of mixing two solid state materials. 4). $SiCl_4$ can decompose quickly and form very fine $SiO_2$ when exposed to the air. This inspired us to use it to apply $SiO_2$ coating to the boron particle surfaces at room temperature.

The amorphous boron powders were mixed with a few drops of $SiCl_4$ (with the appropriate amount calculated to correspond to about 10 wt % Si) so that a very thin layer of $SiO_2$ coated the surface of the B particles. This took place instantly at room temperature, as the $SiCl_4$ decomposed to HCl and formed $SiO_2$. The HCl is highly volatile and can evaporate instantly. This reaction and coating process were carefully carried out in a fume cupboard. The Mg powder was then mixed well with the B powder coated with $SiO_2$. The mixed powders were pelletized and sintered *in-situ* at temperatures in the range of 650-750 °C for just 10 minutes in pure Ar gas. Extra Mg was also added to compensate for the lost of Mg during the sintering. The $MgB_2$ samples made using $SiCl_4$ as silicon source is refereed as $SiCl_4$-$MgB_2$ in this study.

The calculated X-ray diffraction (XRD) patterns using Rietveld refinement fit very well with the observed ones. The refined and observed XRD patterns for the 10 wt% $SiCl_4$ added sample are shown in Fig. 2. The lattice parameters obtained from the refinement revealed that the *a* lattice parameter had been reduced from 3.085 to 3.077 Å for the 10 wt% $SiCl_4$ doped samples as compared to the pure ones. The reduction in the *a* lattice parameter is clearly indicated by the shift of both the (100) and (110) peaks to low diffraction angles, while the (002) peak position remains almost unchanged, as illustrated in the insets in Fig. 2.

It should be noted that the *a* lattice parameter is almost the same as what is commonly seen in so-called C doped $MgB_2$ using various C sources. Fig. 3 displays the reduction in the *a* lattice parameter from some typical $MgB_2$ doped with various C sources[4,14-17] and heat treated at low sintering temperatures, along with the estimated C concentration, and compares them with the $SiCl_4$-$MgB_2$ sample. We can see that all the C doped samples have almost the same *a* value and that, surprisingly, the $SiCl_4$-$MgB_2$ sample shows the largest reduction in the *a* lattice parameter. It is obvious that it would be definitely unreliable to calculate the carbon concentration in C doped $MgB_2$ just from the reduction in the *a* lattice parameter. In other words, reduction of the *a* lattice parameter can not be used as a criterion to judge if the C really is substituted onto boron sites. This lack of correspondence is particularly likely to hold true for samples made at low sintering temperatures with low C doping content and slightly decreased $T_c$. However, for high sintering temperature (close to 1000 $^{o}$C), if both $T_c$ and the *a* lattice parameter are reduced significantly, estimation of the C-content from the *a* lattice parameter should be reasonable.

Another two facts were obtained from the Rietveld refinement: 1). The Mg is very deficient. The occupancies of Mg are about 0.91 and 0.92 for the pure and the $SiCl_4$-$MgB_2$ samples. 2). The *c* lattice parameter for $SiCl_4$-$MgB_2$ is 3.526 Å, slightly greater than for the pure sample (3.524 Å). In order to understand what could cause the changes in the *a* lattice parameter, we carried out a first principles calculation for the following possibilities: 1) Mg deficiency; 2) oxygen occupancy on B sites; 3) both Mg deficiency and oxygen occupancy; and 4) C doping on B sites. We found that none of them can well account for the reduction in the *a* lattice parameter determined from the experimental results, except for the first case of magnesium deficiency. However, as mentioned above, both pure and doped samples have the almost the same level of Mg deficiency. We then take into account another possibility, strain induced effects due to the inclusion of nanoparticles and crystal defects. We found that such strain can indeed further reduce the a lattice parameter. A detailed study of the lattice parameter reduction using first principles calculations will be published elsewhere.[18]

Fig. 4 shows the resistance versus temperature curves (*R-T*) over the temperature range from 30 K to 300 K. The resistivity at 40 K increased from 24 μΩcm for the pure $MgB_2$ to 64 μΩcm for the doped $MgB_2$. The $T_c$ values and residual resistivity ratios, $R(300\ K)/R(T_c)$ (*RRR*), were obtained to be 38.2 and. 36.2 K, and 2.13 and 1.69, for the pure and $SiCl_4$-$MgB_2$ samples, respectively. The reduced RRR values indicate that the scattering is much enhanced in the sample with the addition of $SiCl_4$.

The magnetic field dependence of $J_c$ at 20 and 5 K is shown in Fig. 5. For comparison purposes, the data from an $MgB_2$ sample doped with malic acid ($C_4H_6O_5$), a typical carbon source, is also included in the figure. It is interesting to note that the $J_c$ at 20 K for the $SiCl_4$-$MgB_2$ sample at both low and high fields is higher than for both the pure and the malic acid doped $MgB_2$ samples, which were made under the same conditions as the $SiCl_4$-$MgB_2$. It should be noted that that the $J_c$ of the $SiCl_4$-$MgB_2$ is one order of magnitude higher than for the pure $MgB_2$ at 5 T and the malic acid doped $MgB_2$ at 6 T. The $J_c$ values of the 10 wt% $SiCl_4$ added $MgB_2$ are over 1-2 x $10^4$ A/cm$^2$ at 8 T and 5 K, more than one order of magnitude higher than for the pure $MgB_2$, and almost the same at low fields and comparable to the malic acid doped $MgB_2$ at high fields. These results indicate that the $SiCl_4$ has an the advantage of large $J_c$ at high temperatures over C doped $MgB_2$, as it only decreases the $T_c$ slightly to 36.5 K ,compared to 33 K for the malic acid doped $MgB_2$.

The $H_{c2}$ and $H_{irr}$ were also enhanced in the $SiCl_4$-$MgB_2$, as was proved by the data determined from the R-T curves.. The $H_{c2}$ values versus normalized temperature $T/T_c$ obtained from the 90% or 10% values of their corresponding resistive transitions are shown in Fig. 6. The $H_{c2}$ values of the un-doped sample are also included for comparison. Significantly enhanced $H_{irr}$ and $H_{c2}$ are clearly observed for the doped sample. The above results reveal that the doped $MgB_2$ has higher $H_{irr}$ values compared to the un-doped samples that were processed under the same fabrication conditions.

Now let us consider if the $SiCl_4$ results in the enhancement of flux pinning. We need to know if the enhanced in-field $J_c$ or $H_{irr}$ is caused by the enhancement of $H_{c2}$ or the introduction of more effective pinning centers compared to the un-doped $MgB_2$ samples. As we have explained in the introduction section, we calculated a new ratio, RHH = $H_{c2}/H_{irr}$ for the $SiCl_4$-$MgB_2$ sample. The results are shown in Fig. 7.

As we have discussed earlier, if the RHH value is close to 1, it means that flux pinning is very strong. As can be seen from Fig. 6, the RHH values for the $SiCl_4$ sample are the same as for the un-doped $MgB_2$. This means that the $H_{irr}$ enhancement is not due to the introduction of extra effective pinning centers, instead, its enhancement is a result of the $H_{c2}$ enhancement. We have used this new RHH ratio to evaluate what happened in the C-doped samples with malic acid ($C_4H_6O_5$)[10] and silicone oil doping[11].. We can see that their RHH lines are above those of both pure and $SiCl_4$ doped samples. This clearly demonstrates that the enhancement of $J_c$ in both malic acid and silicone oil doped $MgB_2$ is caused by an increase in $H_{c2}$ rather than flux pinning enhancement. We can also say that the flux pinning in the C-doped $MgB_2$ is "weaker" than in the pure and $SiCl_4$ doped samples. It is interesting to note that the nano-Si doped $MgB_2$ has the same RHH valves as $SiCl_4$, indicating that nano-Si doping[5] enhances $H_{c2}$ without causing extra weak pinning compared to the $MgB_2$ doping with different C sources. The new RHH values or RHH lines give us some guidance on how to further enhance the $J_c$ for $MgB_2$. For C-doping, it is clear that we still have plenty of room to enhance the in-field $J_c$, if we can reduce the RHH values further by optimizing the fabrication conditions.

We now discuss what causes the enhancement of $H_{c2}$ in the $SiCl_4$-$MgB_2$. We first look at the grain sizes, based on the FWHM of the XRD peaks. The diffraction peaks are observed to be broadened, compared with un-doped samples, indicating that the grain sizes are reduced and considerable strain has been introduced.

Fig. 8 (a) shows a TEM image for the $SiCl_4$-$MgB_2$ sample. The sample has an average grain size of several tens of nanometers and contains a high density of defects, such as dislocations and heavy strains in the lattice. Under high magnification (b) we can see a large density of MgO clusters with sizes of 1-3 nm that have been embedded in individual grains.

Based on the facts that the grain sizes are reduced, impurities are in the form of inclusions, and crystal imperfections/defects have inhomogeneous distributions, we expect the enhanced scattering and $H_{c2}$ which have been demonstrated in the previous sections. We now analyze the pinning mechanism in the $SiCl_4$ doped samples.

Two mechanisms of core pinning are predominant in type-II superconductors, i.e., $\delta T_c$ and $\delta l$ pinning[19]. Whereas $\delta T_c$ pinning is caused by the spatial variation of the GL coefficient associated with disorder in the transition temperature $T_c$ , variations in the charge-carrier mean free path $l$ near lattice defects are the main cause of $\delta l$ pinning. The $\delta T_c$ and $\delta l$ pinning mechanisms result in different temperature dependences of $J_{sv}$, where $J_{sv}$ is $J_c$ in the single vortex pinning regime. For $\delta T_c$ pinning, $J_{sv} \propto (1-t^2)^{7/6}(1+t^2)^{5/6}$ with $t = T/T_c$, while for the case of $\delta l$ pinning, $J_{sv} \propto (1-t^2)^{2/5}(1+t^2)^{-1/2}$. Taking into account collective pinning theory[20], $J_c$ is field independent when the applied field is lower than a crossover field $B_{sb}$ at which the dominant pinning mechanism changes from single vortex to small bundle inning. When the single vortex pinning mechanism is dominant $B_{sb} \propto$ the product of $J_{sv}$ and $B_{c2}$. The following $B_{sb}$ temperature dependence can be obtained[21]:

$$B_{sb}(T) = B_{sb}(0)\left(\frac{1-t^2}{1+t^2}\right)^{\upsilon}$$

where $\upsilon$= 2/3 and 2 for $\delta T_c$ and $\delta l$ pinning, respectively.

The curves of $B_{sb}$(T) for $\delta T_c$ and $\delta l$ pinning is shown in Fig. 9. The curve has a positive curvature for $\delta T_c$ pinning, while the curvature associated with $\delta l$ pinning is negative. It is clear that the $B_{sb}$(T) showing a negative curvature with $\upsilon$ = 2 is in good agreement with the experimental data. This strongly suggested that $\delta l$ pinning is the dominant pinning mechanism in our $SiCl_4$-$MgB_2$ sample. It is believed that the nano-MgO inclusions with variable distances and sizes (1-10 nm ) shown in Fig 7(b) are the main type of defect and the reason for the variation in the charge-carrier mean free path *l*.

Acknowledgement

This work is supported by the Australian Research Council.

Figure and figure captions

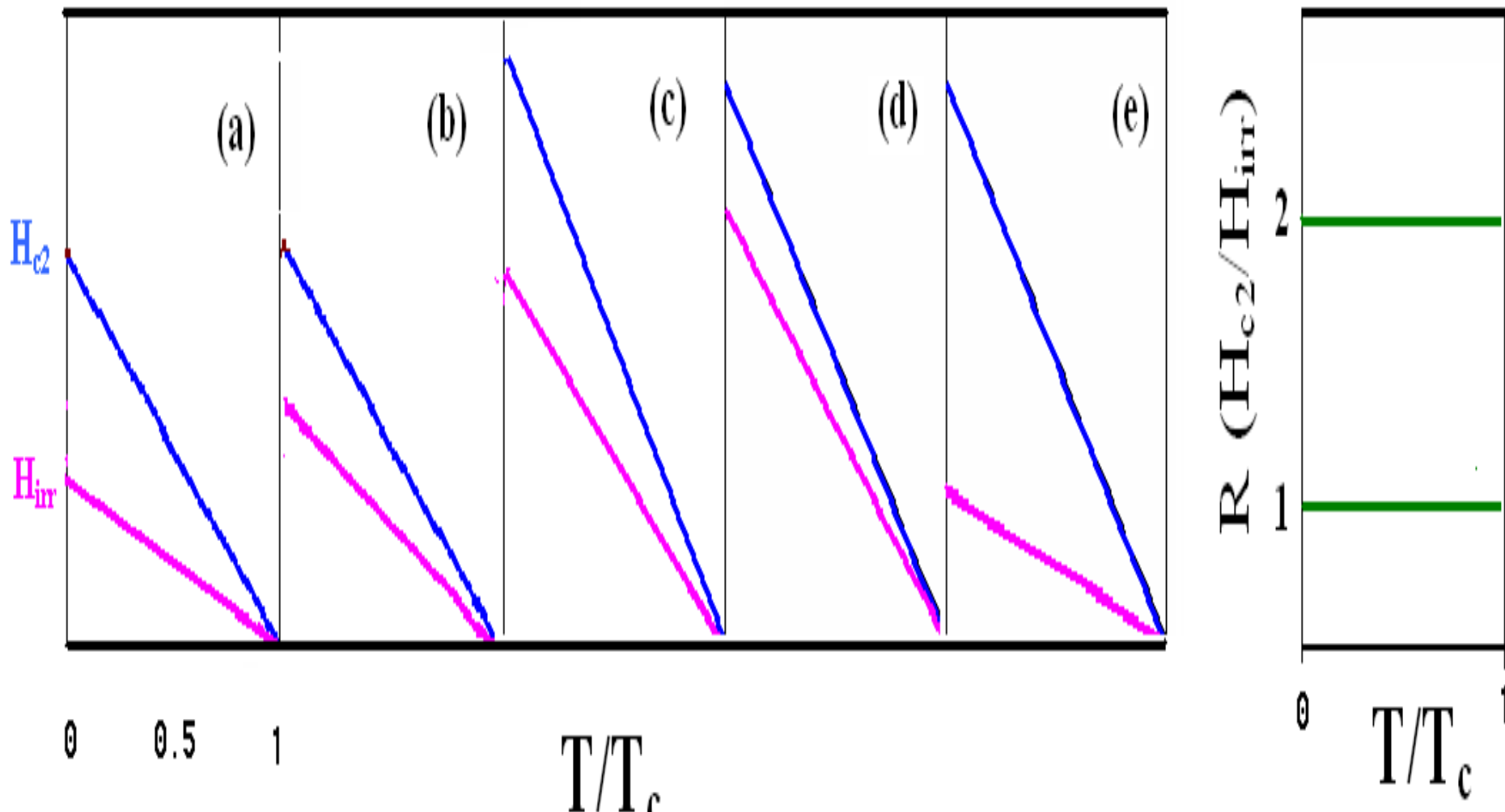

Figure 1. Schematic showing magnetic fields of $H_{v2}$ and $H_{irr}$ lines as a function of reduced temperature $T/T_c$ for different types of samples (a) – (e). RHH ratio as a function of reduced temperature (f), showing cases where the $H_{irr}$ and $H_{c2}$ lines coincide (RHH = 1) and where there is fairly weak pinning (RHH = 2). For samples with good pinning performance, the RHH valves should be located close to 1.

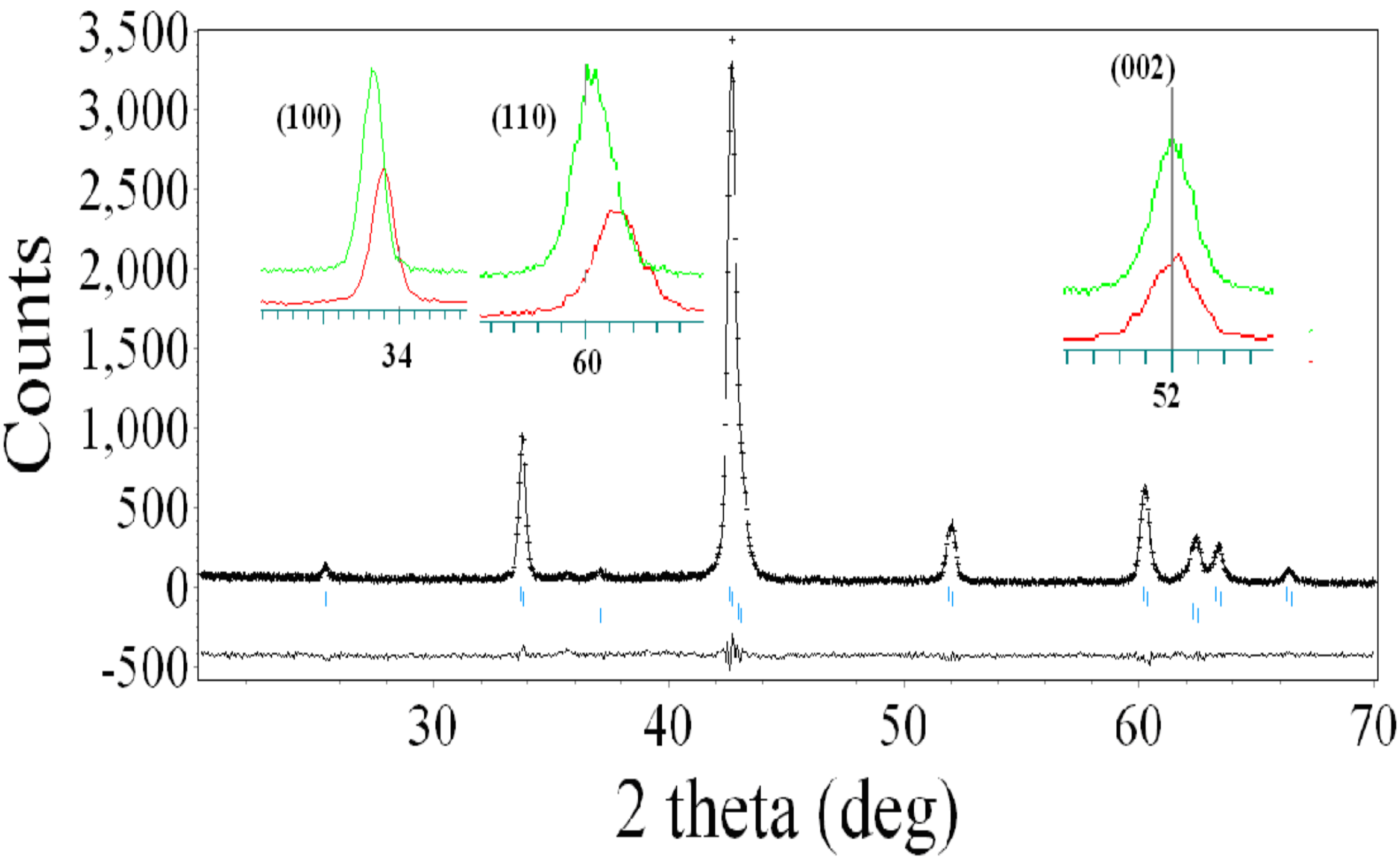

Figure 2. XRD pattern of the observed (crosses), calculated (solid line), and difference (bottom solid line) diffraction profiles at 300K for $MgB_2$ with 10 wt% $SiCl_4$ added. The upper and lower peak markers relate to $MgB_2$ and MgO, respectively. The insets show key XRD peaks for the doped (green) and the pure (red) samples.

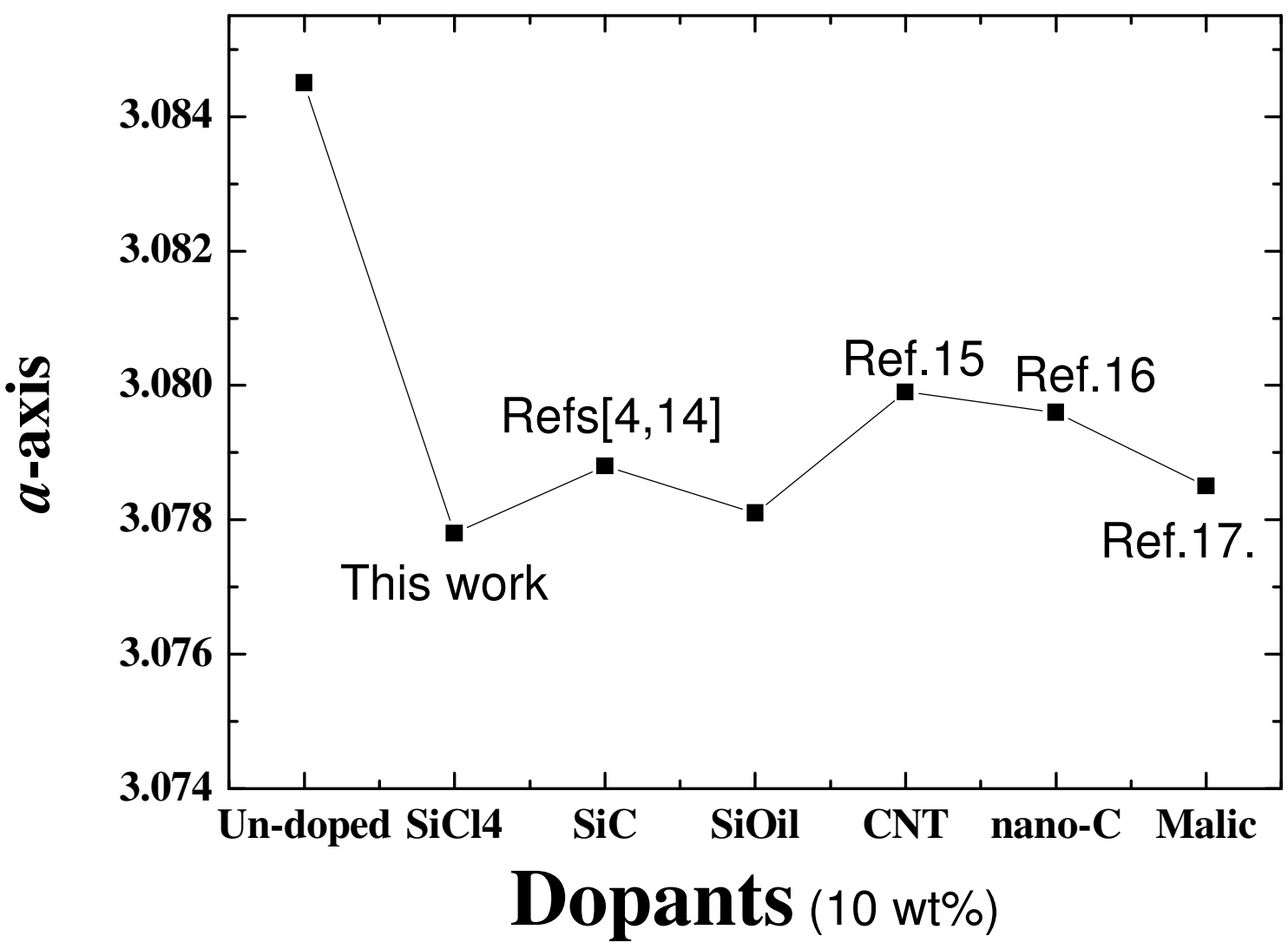

Figure 3. a-axis lattice parameter in various doped and un-doped $MgB_2$ samples.

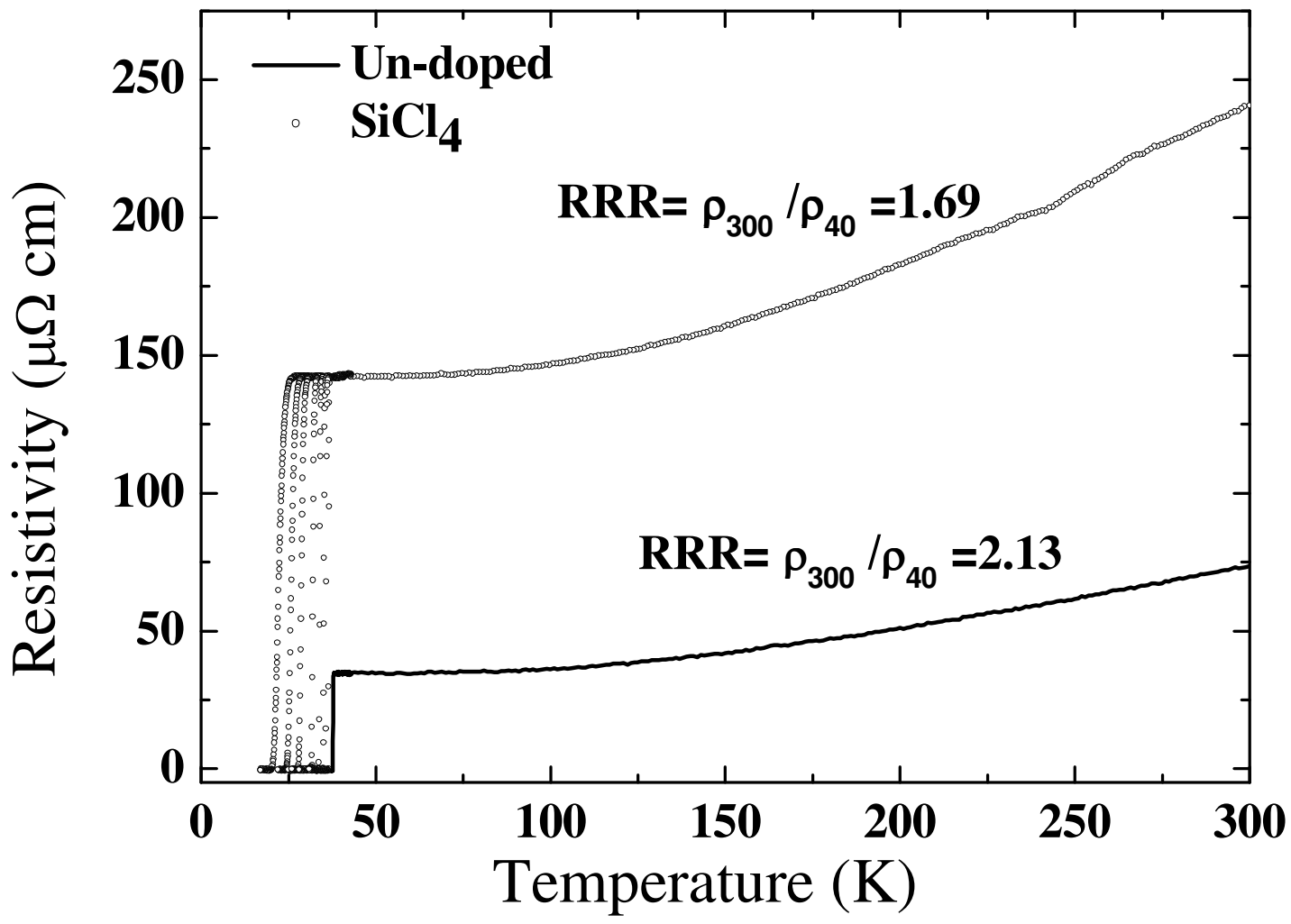

Figure 4. Temperature dependence of the resistivity measured at fields up to 8.7 T for 10 wt% $SiCl_4$ doped and pure $MgB_2$**.**

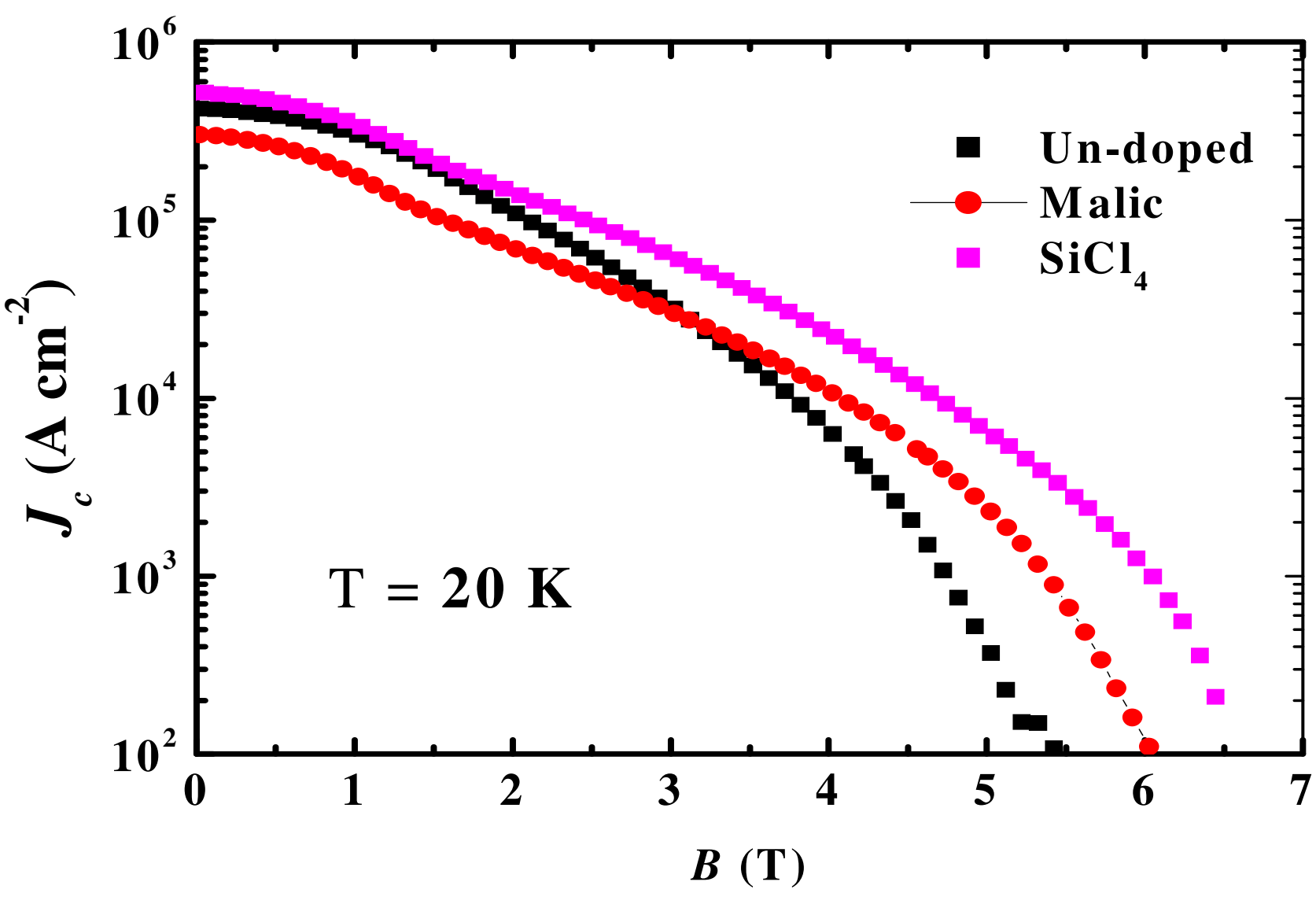

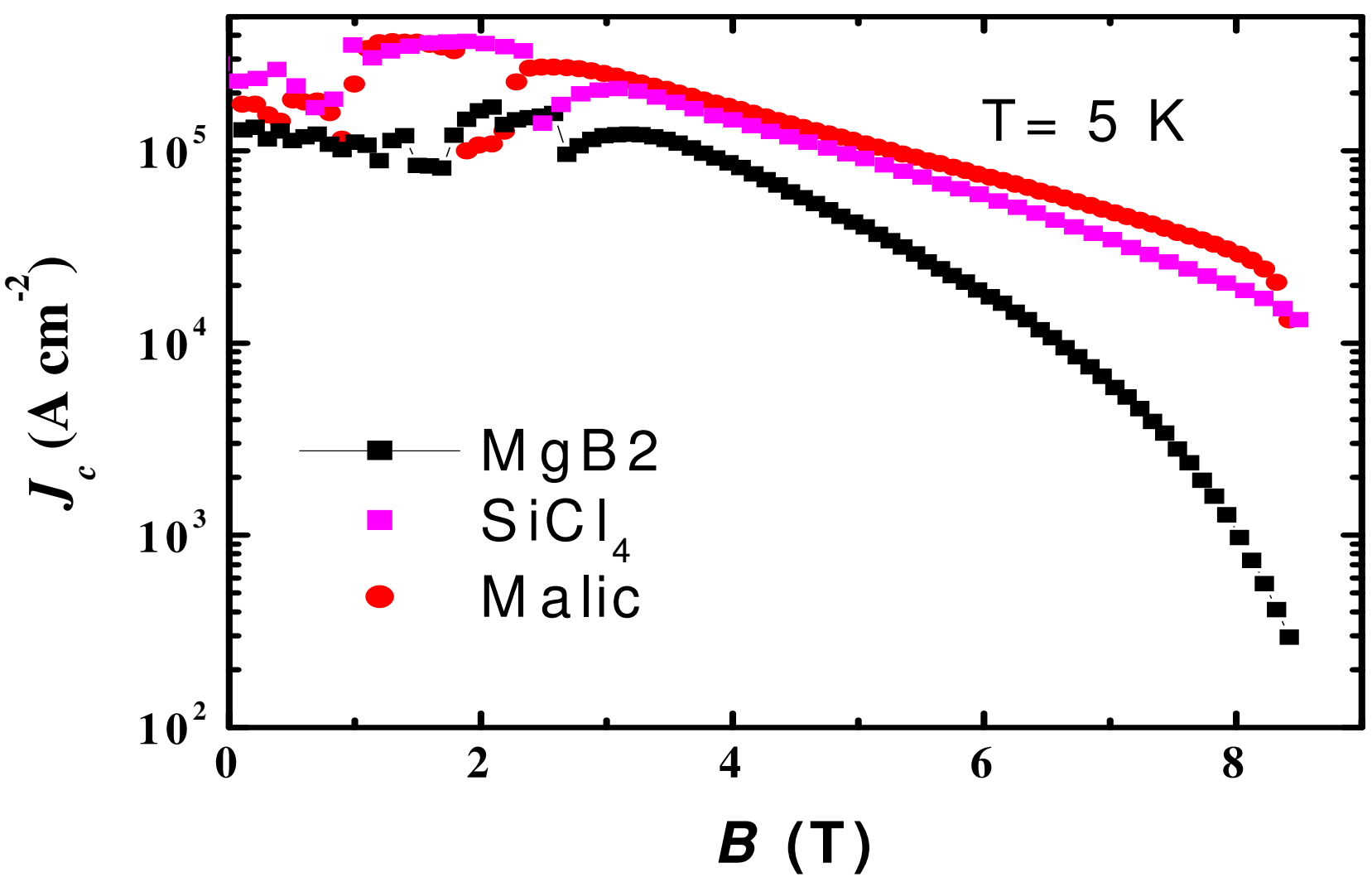

Figure 5. Magnetic field dependence of the critical current at 20 K (a) and 5 K (b) for pure, 10 wt% $SiCl_4$ doped, and malic acid doped $MgB_2$.

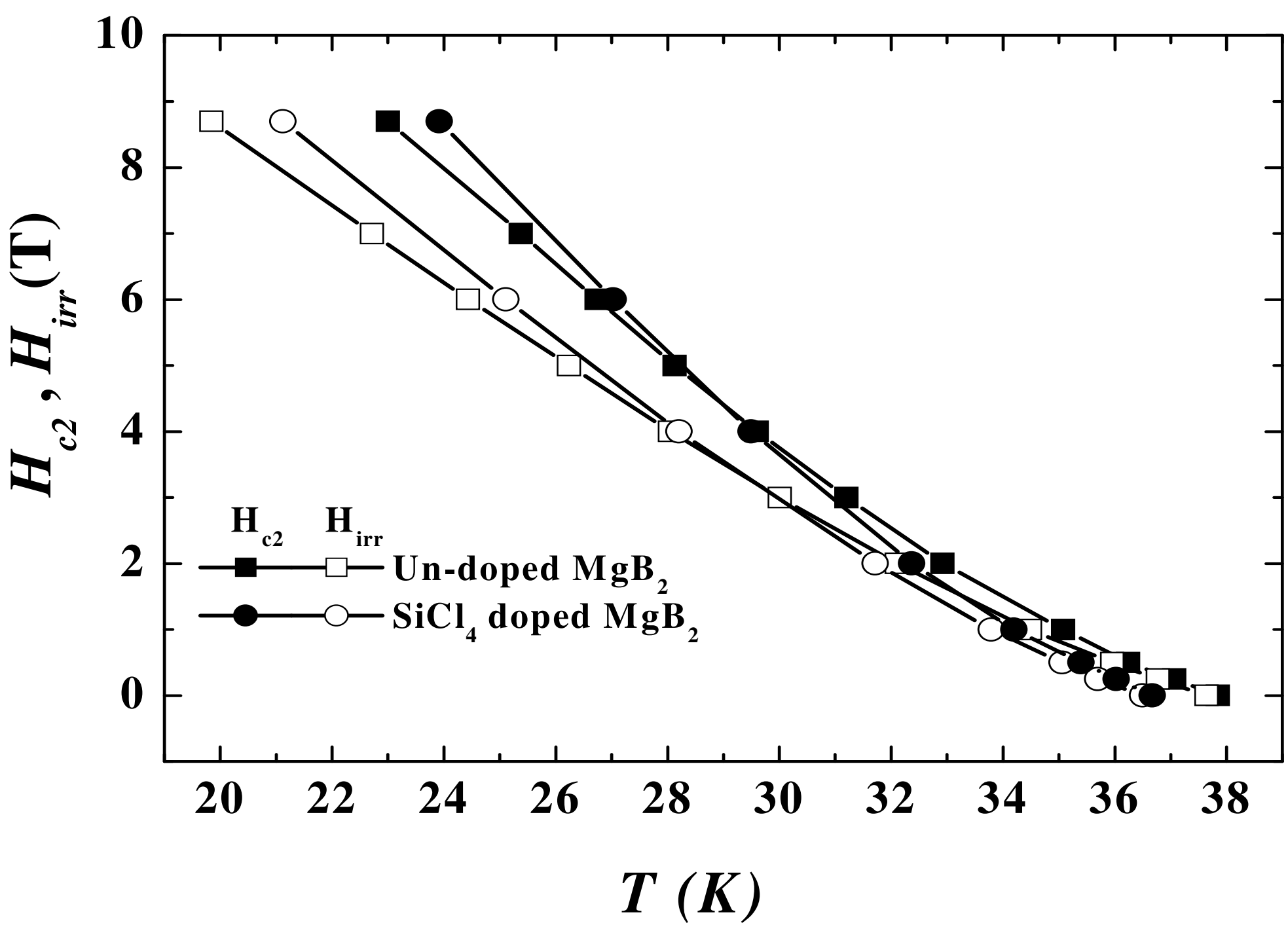

Figure 6. Temperature dependence of $H_{irr}$ and $H_{c2}$ for pure $MgB_2$ and $SiCl_4$-$MgB_2$.

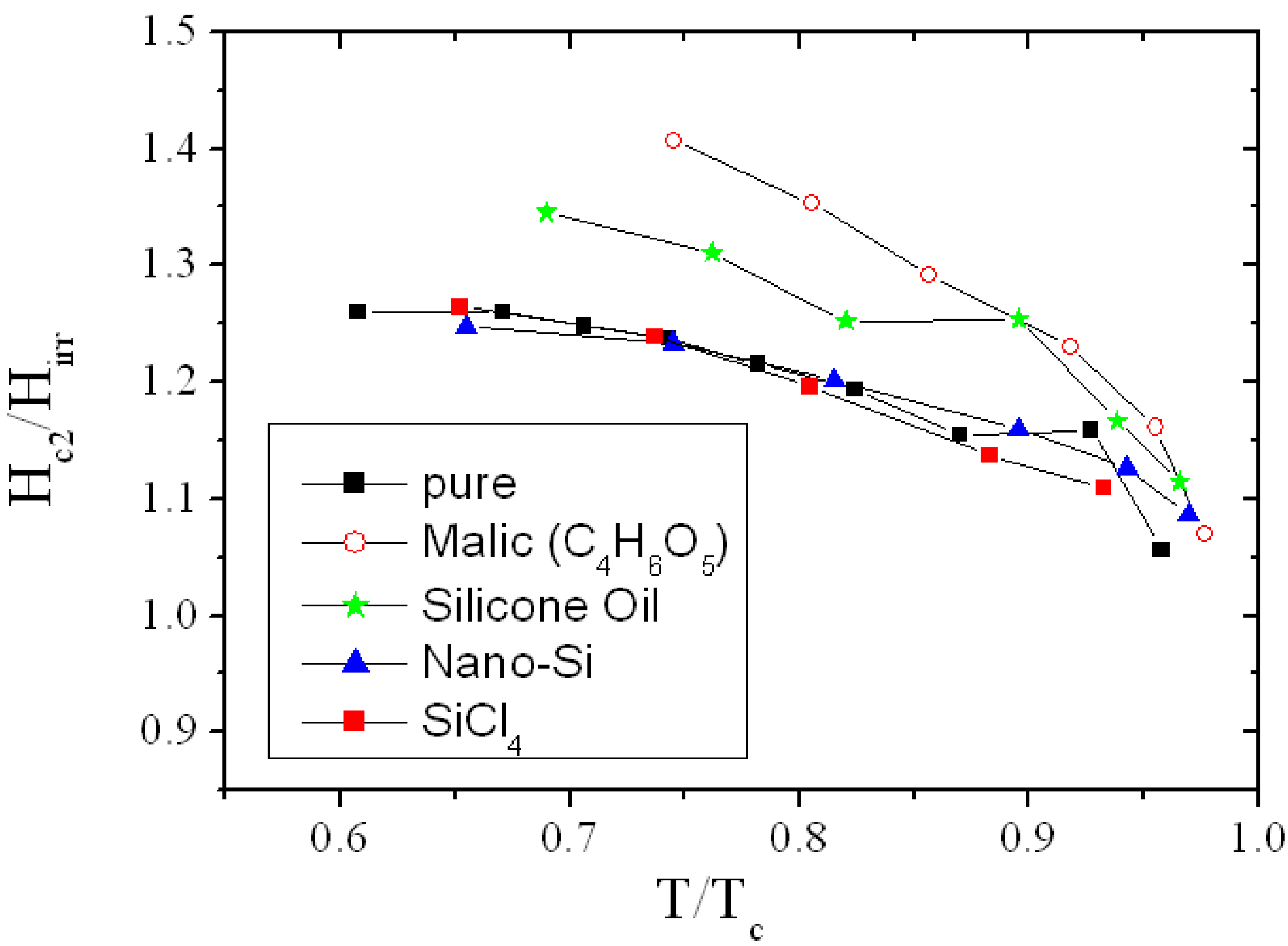

Figure 7. RHH ratio as a function of reduced temperature for various $MgB_2$ samples doped with 10 wt% chemicals and sintered at the same temperature.

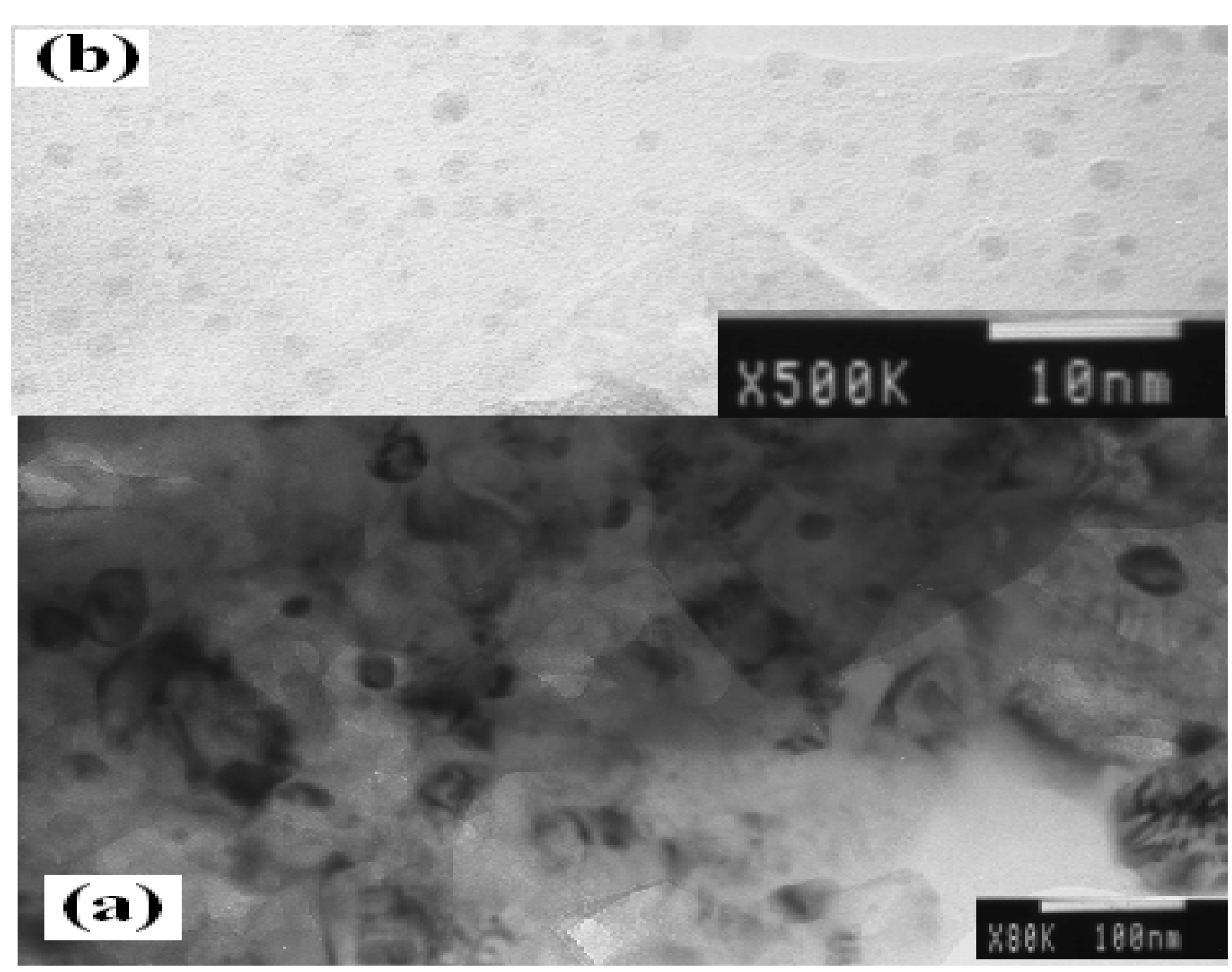

Figure 8. TEM images of 10 wt% $SiCl_4$ doped $MgB_2$.

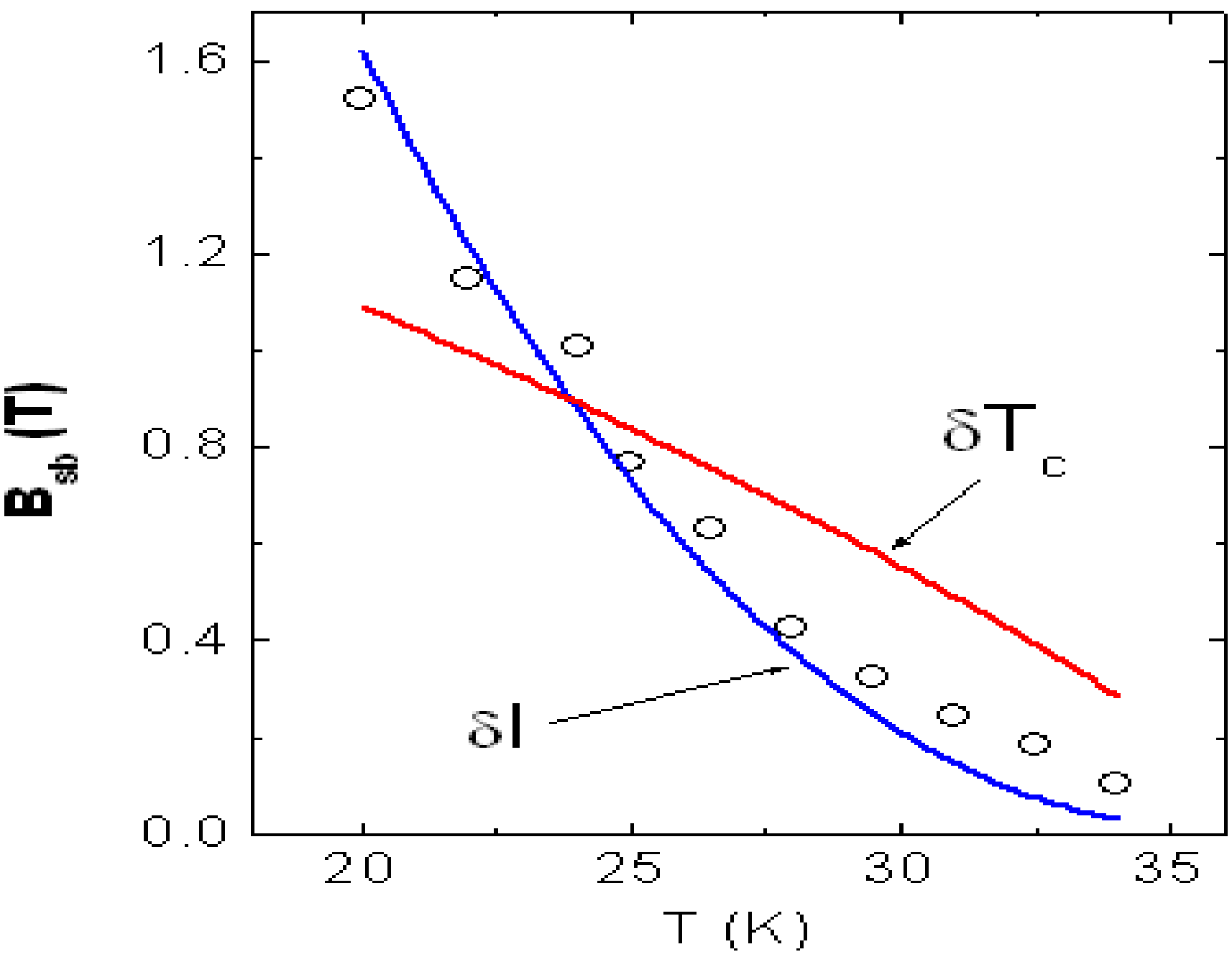

Figure 9. Temperature dependence of the crossover field $B_{sb}$ for the 10 wt% $SiCl_4$ doped sample, , the $\delta l$ pinning curve with $\upsilon = 2$ and the $\delta T_c$ pinning curve with $\upsilon=3/2$.